\documentclass[sigconf, nonacm, pdfa]{acmart}
\usepackage[a-2b]{pdfx}

\usepackage[utf8]{inputenc}
\usepackage{tabularx}
\usepackage{graphicx}
\usepackage{booktabs}
\usepackage{threeparttable}
\usepackage{amsmath}
\usepackage{makecell} 
\usepackage{listings}
\usepackage{xcolor}
\usepackage{subcaption}

\definecolor{codeblue}{rgb}{0,0,0.8}
\definecolor{codegreen}{rgb}{0,0.5,0}
\definecolor{codegray}{rgb}{0.4,0.4,0.4}

\lstdefinestyle{cleanpython}{
    backgroundcolor=\color{white},   
    commentstyle=\color{codegreen},
    keywordstyle=\color{codeblue},
    numberstyle=\tiny\color{codegray},
    stringstyle=\color{purple},
    basicstyle=\ttfamily\small,
    breaklines=true,                 
    captionpos=b,                    
    numbers=left,                    
    numbersep=10pt,                  
    showstringspaces=false,
    frame=single,                    
    rulecolor=\color{lightgray},     
    language=Python
}

\newcommand\vldbdoi{XX.XX/XXX.XX}

\newcommand\vldbvolume{14}
\newcommand\vldbissue{1}
\newcommand\vldbyear{2026}
\newcommand\vldbauthors{\authors}
\newcommand\vldbtitle{\shorttitle} 
\newcommand\vldbavailabilityurl{URL_TO_YOUR_ARTIFACTS}
\newcommand\vldbpagestyle{plain}

\newcommand\vldbworkshop{Applied AI for Database Systems and Applications (AIDB 2026)}

\begin{document}

\title{AI Query Compilation for Unified and Optimized Execution}

\author{Yeounoh Chung}
\authornote{Corresponding author.}
\affiliation{%
  \institution{Google Cloud}
  \city{Sunnyvale}
  \state{CA}
  \country{USA}
}
\email{yeounoh@google.com}

\author{Helena Caminal}
\affiliation{%
  \institution{Google Cloud}
  \city{Sunnyvale}
  \state{CA}
  \country{USA}
}
\email{hcaminal@google.com}

\author{Fatma \"{O}zcan}
\affiliation{%
  \institution{Google Cloud}
  \city{Sunnyvale}
  \state{CA}
  \country{USA}
}
\email{fozcan@google.com}

\renewcommand{\shortauthors}{Chung et al.}

\begin{abstract}
In this vision paper, we propose a novel architectural paradigm for accelerated AI query execution via a unified compiled execution strategy. By compiling the hybrid \textit{AI Query} as a whole—integrating both standard SQL relational constructs and LLM inference layers into a single, unified tensor compute graph—we completely alleviate PCIe data movement bottlenecks across execution boundaries and enable global compiler optimizations and efficient automatic sharding. We demonstrate the viability of this unified execution paradigm on select and extended AI queries on \textit{SemBench} Reviews and Movies datasets, achieving up to 5.3x latency speedup and 9.8x throughput speedup on TPUs, and outline a research roadmap of open technical challenges to realize this vision.
\end{abstract}

\maketitle

\pagestyle{\vldbpagestyle}
\begingroup\small\noindent\raggedright\textbf{VLDB Workshop Reference Format:}\\
\vldbauthors. \vldbtitle. VLDB \vldbyear\ Workshop: \vldbworkshop.\\ 
\endgroup
\begingroup
\renewcommand\thefootnote{}\footnote{\noindent
This work is licensed under the Creative Commons BY-NC-ND 4.0 International License. Visit \url{https://creativecommons.org/licenses/by-nc-nd/4.0/} to view a copy of this license. For any use beyond those covered by this license, obtain permission by emailing \href{mailto:info@vldb.org}{info@vldb.org}. Copyright is held by the owner/author(s). Publication rights licensed to the VLDB Endowment. \\
\raggedright Proceedings of the VLDB Endowment, Vol. \vldbvolume, No. \vldbissue\ %
ISSN 2150-8097. \\
}\addtocounter{footnote}{-1}\endgroup

\ifdefempty{\vldbavailabilityurl}{}{
\vspace{.3cm}
\begingroup\small\noindent\raggedright\textbf{PVLDB Artifact Availability:}\\
N/A
\endgroup
}

\section{Introduction}

Relational query processing through hardware accelerators, such as GPUs and TPUs, has been an active area of research~\cite{he2022query,hu2022tcudb,holanda2019relational,rosenfeld2022query,zhang2013omnidb,wang2014concurrent,volk2010gpu,chrysogelos2019hardware,mageirakos2026togpu,li2025scaling}. However, when it comes to AI queries or SQL queries augmented with AI semantic operators (e.g., using an LLM for semantic classification or extraction)~\cite{chung2026100x,liskowski2026cortex,patel2025semantic,liu2024declarative}, current systems typically adopt a split execution model. Traditional SQL operations are executed on the CPU or a separate database engine, while LLM inference is offloaded to a dedicated ML serving system, often running on a remote machine or a separate accelerator. 

Unfortunately, this dichotomy creates a two-fold penalty that limits the performance of modern AI queries. First, it imposes substantial data movement overheads, as intermediate results must be transported across the Host-To-Device (H2D) boundaries between the database engine's system DRAM and the inference engine's high-bandwidth memory (HBM) or network-attached remote computing resources such as Google Vertex AI~\cite{cloud_ai_platforms}. This fragmentation exacerbates the classic \textit{Memory Wall}~\cite{wulf1995hitting}, where the performance of the system is dictated not by raw compute capability, but by the bandwidth and latency of data movement~\cite{gholami2024ai}. In the AI era, this has evolved into a \textit{Data Movement Wall}~\cite{huang2026cidr}, where the cost of marshaling data between isolated execution environments outweighs the gains of specialized hardware acceleration.
Second, these rigid execution boundaries can also act as an optimization huddle, precluding any form of global AI query planning. Because the SQL optimizer remains blind to the downstream neural operations, while the ML compiler also remains unaware of the upstream relational context, potential cross-boundary optimizations (e.g., SQL and AI operator fusion) are overlooked.

In this paper, we propose a paradigm shift toward a unified  execution strategy for AI queries. Our key insight is that since modern LLM inference models are fundamentally executed as tensor compute graphs, we can bridge the gap by lowering standard relational database operations (e.g., scans, filters and aggregations) into that same tensor compute graphs. By unifying relational constructs and neural layers into a consolidated executable, we can leverage the full power of modern ML compiler infra, such as JAX~\cite{frostig2019compiling} or MLIR~\cite{lattner2021mlir}.
This allows the entire hybrid flow to be compiled into a single accelerator executable, optimized holistically for specialized hardware backends, like TPUs, GPUs or other specialized hardware backends for MLIR. 

This unified approach provides more than just the data movement latency elimination; it fundamentally removes the optimization barriers due to the fragmented execution environments, unlocking new opportunities for global cross-boundary optimizations for AI queries or any analytics workloads that implement both traditional SQL and AI semantic operators.
A major global optimization enabled by this unified program is unified, compiler-driven multi-device sharding. In traditional split AI Query Engine architectures, developers must manually write, calibrate, and synchronize two completely independent parallelization strategies: one for the CPU database engine (e.g., dynamic core-level partitioning) and another for the TPU/GPU machine learning engine (e.g., tensor-parallel or pipeline-parallel model sharding). This requires complex dynamic load balancing across the CPU-TPU boundary. By contrast, because our compiler translates both relational database scans and LLM inferences into a single unified tensor program, the compiler can globally shard both the relational data arrays and the neural computations concurrently and automatically via XLA's GSPMD/PMAP. This completely eliminates the need to write separate parallelization logic for the CPU and TPU engines, allowing the compiler to dynamically and optimally partition the entire hybrid query end-to-end across hardware resources.

\section{Related Works}
Our work is at the intersection of relational database hardware acceleration and LLM-based AI Query engine optimizations.

\noindent\textbf{Relational Algebra on Accelerators:} 
accelerating SQL query processing on specialized hardware accelerators (e.g., GPUs and TPUs) has been an active area of research. Early works in tensor query processing mapped standard relational algebra operators (e.g., selections, joins and projections) to GPU tensor kernels~\cite{he2022query,hu2022tcudb}. Recent research projects explore leveraging accelerator hardwares for graph databases~\cite{li2024tengraph} and other broader ML workflows~\cite{boehm2023optimizing}.

Our work is orthogonal to these relational query acceleration works. While they focus entirely on enabling highly optimized relational query processing on hardware accelerators, our work specifically targets AI queries or hybrid queries (SQL + LLMs). Our contribution is showing how to unify relational algebra and deep neural layers into the same compiled tensor program, completely consolidating the accelerator execution boundaries and their cross-boundary data movements. The unified compiled execution via JAX/XLA also enables auto-sharding, which otherwise requires separate parallelization and optimizations on CPU database engines and the TPU/GPU ML serving frameworks.

\noindent\textbf{AI Query Systems \& Serving Boundaries:} 
as LLMs have grown in popularity, several contemporary data analytics systems have integrated LLM-based semantic operators into relational query engines for structured and unstructured data analytics~\cite{chung2026100x,liskowski2026cortex,patel2025semantic}. These systems universally adopt a split execution model: traditional SQL or other data processing pipelines run eagerly on the CPU database host, whereas semantic operators (LLM inference) is offloaded to a local or remote model serving engine (e.g., vLLM~\cite{kwon2023efficient}) running on GPUs/TPUs.
While recent industry architectures envision unified, AI-powered agentic data clouds where AI agents and database operators run seamlessly on consolidated enterprise platforms~\cite{chung2026architecting}, contemporary physical implementations still strictly separate relational engines from LLM serving boundaries.

\section{The Case for Unified AI Query Execution}
Our system achieves accelerated AI Query execution via a compiled unified execution path: compiling the AI Query as a whole, integrating both standard SQL relational constructs and the deep learning neural layers (e.g., LLM inference for semantic operators) into a single, unified compute graph. Contemporary AI Query engines operate as a fragmented mixture execution model, split by fixed hardware boundaries. Standard relational operations run eagerly on CPU database engines, whereas LLM inferences are offloaded to separate ML serving frameworks on accelerators (GPUs/TPUs). This split-boundary execution incurs severe data movement overheads and prevents global compiler optimizations.
We translate both relational operators and LLM inference into a unified JAX program, compiled into a single global TPU/GPU executable via XLA JIT~\cite{frostig2019compiling}. This consolidated approach is illustrated in Figure~\ref{fig:aiq_overview_small}.

\begin{figure}[t]
  \centering
  \includegraphics[width=\columnwidth]{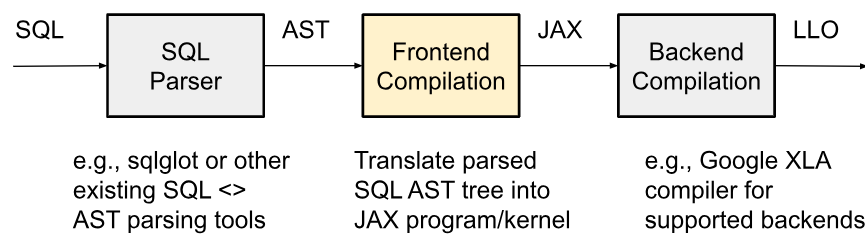}
  \caption{High-level view of the unified AI Query compilation flow.}
  \label{fig:aiq_overview_small}
\end{figure}

\subsection{Translation and Mapping Mechanics}
\label{sec:frontend_compilation}
\begin{table*}[!ht]
    \centering
    \small
    \renewcommand{\arraystretch}{1.5}
    \begin{tabularx}{\textwidth}{|l|l|l|X|}
    \hline
    \textbf{SQL Clause / Operator} & \textbf{AST Node (\texttt{sqlglot})} & \textbf{Physical Translation} & \textbf{Compiled JAX Implementation} \\ \hline
    \texttt{FROM table} & \texttt{exp.From} & Input Array Loading & Standard JAX input arrays packed into host sharded memory (\texttt{input\_ids}, \texttt{attn\_mask}, \texttt{values\_a}). \\ \hline
    \texttt{WHERE values\_a < 0.2} & \texttt{exp.Where} & Boolean Masking & Vectorized JAX boolean array mask generated dynamically: \texttt{values\_a < 0.2} (evaluates to \texttt{True} / \texttt{False} array). \\ \hline
    \texttt{AI\_IF(text\_col) = 1} & \texttt{exp.Func} (WHERE) & LLM Classification & High-level UDF calling Flax causal LM forward pass. Logits for ``Yes''/``No'' are compared on-chip to yield boolean predicates. \\ \hline
    \texttt{AI\_TRANSFORM(text\_col)} & \texttt{exp.Func} (SELECT) & LLM Generation & Flax causal LM forward pass with generative decoding, output sharded and masked via \texttt{jnp.where} and \texttt{pad\_id}. \\ \hline
    \texttt{SELECT SUM(values\_a)} & \texttt{exp.Agg} & Masked Reduction & Vectorized summation following the vectorized JAX boolean array masking: \texttt{jnp.where(final\_predicate, values\_a, 0.0).sum()}. \\ \hline
    \end{tabularx}
    \caption{SQL to JAX translation mapping examples}
    \label{tab:jax_mapping}
\end{table*}

Each component of a hybrid SQL query maps to SQL AST constructs, which our compiler translates directly to standard JAX functions/primitives. Table~\ref{tab:jax_mapping} describes the structural mapping mechanics from SQL clauses to their compiled JAX implementations.

As a running example, let us consider a hybrid AI Query that filters rows using a semantic predicate evaluated by an LLM:
\texttt{SELECT SUM(values\_a) FROM bench\_table WHERE AI\_IF(text\_col, `Is this review positive?') = 1}.

To compile this AI Query with \textit{AI\_IF} semantic classification AI operator, the system first parses the SQL text into an Abstract Syntax Tree (AST). The AST tree is subesquently mapped and compiled into a single JAX program (Figure~\ref{fig:jax_kernel} per our running example).
\begin{figure}[htbp]
    \centering
\begin{lstlisting}[style=cleanpython]
@jax.jit
def classification_query_kernel(params, input_ids, attn_mask, values, yes_id, no_id):
    """
    SELECT SUM(values) WHERE Model_Predicts(Next_Token) == 'Yes'
    """
    
    # Logits shape: (Batch_Size, Sequence_Length, Vocab_Size)
    outputs = model(input_ids=input_ids, attention_mask=attn_mask, params=params)
    
    # Slice [:, -1, :] to get the predictions at the very end of the sequence.
    last_token_logits = outputs.logits[:, -1, :]
    
    # Look up the specific columns for Yes and No
    score_yes = last_token_logits[:, yes_id]
    score_no = last_token_logits[:, no_id]
    filter = score_yes > score_no # Filter
    
    # Aggregation
    return jnp.where(filter, values, 0.0).sum()
\end{lstlisting}
    \caption{Actual generated JAX program containing the vectorized execution trace for the example AI Query. This is compiled via \textit{@jax.jit} using XLA for optimized hardware execution.}
    \label{fig:jax_kernel}
\end{figure}

\subsection{Frontend Code Compilation}
Our system performs frontend compilation as to generate JAX program/kernel for XLA backend compilation. The frontend code generator, if you will, reads the parsed SQL AST and traverses it to extract relational predicates, UDF (AI operators) names, and table definitions. Based on the translation rules in Table~\ref{tab:jax_mapping}, the compiler retrieves matching template files written in JAX/Python from the UDF (AI operators) registry and inserts them into the global JAX program/function template. Figure~\ref{fig:aiq_overview_small} illustrates the code generation flow.

\begin{figure}[t]
  \centering
  \includegraphics[width=\columnwidth]{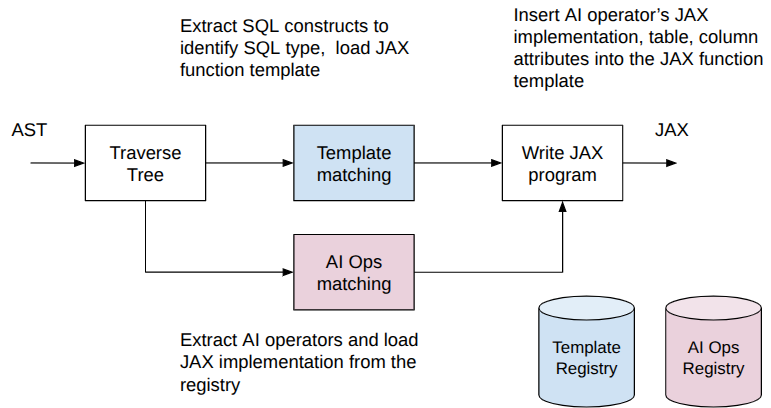}
  \caption{Frontend compilation flow from SQL AST to unified JAX program.}
  \label{fig:aiq_overview_small}
\end{figure}

Note that JAX provides high-performance control flow operator, such as \textit{jax.lax.scan}, which expresses the loop structure (e.g., for, while) without unrolling into a huge static program. This is crucial for memory management, as the generated scan loop streams batches of data through the LLM forward pass directly on the TPU, maintaining the intermedidate states in TPU registers.

\subsection{H2D Boundary Consolidation}
By compiling both standard relational operations (e.g., \textit{WHERE} filters and \textit{SUM} aggregations) and LLM UDFs into the same compiled trace, we achieve host to accelerator device (H2D) boundary consolidation.
In the baseline, mixture execution model, data is moved back and forth between the CPU database engine and the TPU model serving engine:
\begin{itemize}
\item \textbf{H2D (Host-to-Device)} copy of the filtered database rows into TPU HBM memory.
\item \textbf{LLM Inference} model inference for semantic operators executed on the TPU
\item \textbf{D2H (Device-to-Host)} copy of the large output logits back to CPU Host RAM
\item \textbf{Host Processing} data processing and aggregation, like \textit{SUM}, executed on CPU.
\end{itemize}

The D2H copy of output logits can be expensive. For a typical batch size of 128 and sequence length of 32, and Gemma-2B (vocabulary size of 256,128 tokens/bytes) can yield 1GB of output logits per batch that must be copied across the PCIe bus to CPU RAM.

Our unified execution strategy consolidates this boundary; the relational filter predicate is evaluated on the TPU, as well as the LLM forward pass (semantic classsificaiton) and the final \textit{SUM} aggregation/reduction. Consequently, no logits or intermediate arrays are copied back to the host. Only the final single scalar sum or the final generated tokens are copied back to CPU RAM, reducing D2H transfer data sizes from gigabytes to kilobytes. This consolidation completely removes PCIe bug bottlenecks and enables the massive throughput speedups demonstrated in Section~\ref{sec:eval_test}.

\section{Preliminary Results}
We evaluated our AI query compilation prototype on a Google TPUVM with four TPU v5e~\cite{google_cloud_tpu_v5e}) devices. For the initial benchmark, we processed a dataset of 49,664 rows to measure performance at scale and observe horizontal multi-device scaling behaviors. 

\begin{table*}[htbp]
\centering
\renewcommand{\arraystretch}{1.5}
\begin{tabular}{|c|p{0.3\linewidth}|p{0.6\linewidth}|}
\hline
\textbf{} & \textbf{Description} & \textbf{AI Query} \\
\hline
Q1 & Aggregation with semantic filtering & \texttt{SELECT SUM(values\_a) FROM movie\_table WHERE AI\_IF(reviewText, `Is this review positive?') = 1} \\
\hline
Q2 & Aggregation with relational and semantic filtering & \texttt{SELECT SUM(values\_a) FROM movie\_table WHERE values\_a > 70.0 AND AI\_IF(reviewText, `Is this review positive?') = 1} \\
\hline
Q3 & Semantic transformation with relational predicate & \texttt{SELECT AI\_TRANSFORM(reviewText, `One sentence summary of review:') FROM movie\_table WHERE values\_a < 40.0} \\
\hline
Q4 & Classification with cross join & \texttt{SELECT A.reviewText, B.reviewText, AI\_IF([A.reviewText, B.reviewText], `Almost identical reviews?') FROM movie\_table A CROSS JOIN movie\_table B WHERE values\_a < 40.0 AND A.id = B.id AND A.review\_id < B.review\_id} \\
\hline
Q5 & Aggregation across long output sequences with extra debug outputs & \texttt{SELECT COUNT(*) FROM movie\_table WHERE AI\_TRANSFORM(reviewText, 'Rewrite this review as a detailed 3-page critical essay', debug=True) LIKE ``\%masterpiece\%"} \\
\hline
\end{tabular}
\caption{Distinct AI queries. Q1-2 (aggregation with semantic filtering) are taken from \textit{SemBench} for Rotten Tomatoes Critic Reviews and Movies datasets; Q3-4 are extended/synthesized AI queries for more rigorous testing; Q5 is an aggregation query with long output sequences and also with various debug information (e.g., full sequence logits to explain the LLM outputs)}
\label{tab:ai_sql_queries}
\end{table*}

\noindent\textbf{Baseline and Evaluation Scope:} For our baseline, we compare our unified compiled program against an eager mixture execution model (CPU for standard SQL and TPU for LLM inference). This baseline represents the standard architecture of contemporary AI Query Engines~\cite{chung2026100x,liskowski2026cortex,patel2025semantic,liu2024declarative}, where standard relational query steps and data processing runs on the CPU host DBMS and the semantic operators/UDFs are run on hardware accelerators outside DBMS or even offloaded to an independent ML serving framework.

It is important to note that optimizing traditional SQL processing performance on the TPU hardware is not within the scope of this paper. Our compiler does not write customized, low-level TPU hardware kernels nor does it seek to build a highly optimized database engine on TPUs. Traditional SQL processing optimization on TPUs (or GPUs) is a separate, interesting area of research in itself, and our design is orthogonal to those efforts~\cite{he2022query,holanda2019relational,rosenfeld2022query,zhang2013omnidb,wang2014concurrent,volk2010gpu,chrysogelos2019hardware}.

Having said that, our evaluation scope is strictly focused on demonstrating the feasibility, scalability and benefits of extra data-copy elimination via the unified execution path for hybrid AI queries. Our primary contribution is showing that we can compile a single program for such hybrid AI queries, which effectively eliminate the cross-boundary data movements and yield more performance optimization opportunities. Furthermore, by compiling the entire hybrid query as a single program, we make multi-device horizontal scaling fully automatic and easy via XLA's GSPMD/\textit{pmap}, sharding both relational arrays and model weights. Consequently, we do not compare our prototype system directly with highly customized GPU-based database systems. 

\noindent\textbf{Evaluation Queries and Datasets:} 
To evaluate our approach on a more realistic AI query workload, we used a select analytics queries from \textit{SemBench}~\cite{lao2026sembenchbenchmarksemanticquery}. In particular, we evaluate the three AI queries using the Rotten Tomatoes Critic Reviews and Movies datasets -- the reviews table is joined with the movie details table on movie \textit{id}, and we use textit{audienceScore} as a traditional filter column \textit{values\_a}. We originally focused on the Movies scenario with text-only semantic filtering (\textit{AI\_IF}). To evaluate our approach across a broader suite of operators, we synthesized a semantic mapping query (\textit{AI\_TRANSFORM}) on top of this text workload. While the benchmark includes out-of-the-box semantic mapping queries in other scenarios, those native workloads are heavily multi-modal (e.g., joining textual attributes with images). This would pose an orthogonal compiler complexities (e.g., using VLM~\cite{bordes2024introduction}), which is beyond the scope of this paper and our preliminary evaluation.
We compiled and evaluated five distinct AI queries from Table~\ref{tab:ai_sql_queries}, at a scale of up to 49664 reviews.




\subsection{Movie Review Analysis Performances}\label{sec:sembench}
We evaluate aggregation and transformation AI queries (Queries 1-3 in Table~\ref{tab:ai_sql_queries}) based on \textit{SemBench}, that represent more realistic AI query workloads. 

\begin{table}[ht]
    \centering
    \caption{SemBench Movie workload performance comparison on 4 TPU devices.}
    \label{tab:sembench_performance}
    \resizebox{\columnwidth}{!}{%
        \begin{tabular}{@{}lccccc@{}}
            \toprule
            \makecell[l]{\textbf{Query Type}} & 
            \makecell{\textbf{Baseline Latency}\\\textbf{(ms)}} & 
            \makecell{\textbf{Unified}\\\textbf{Latency (ms)}} & 
            \makecell{\textbf{Throughput}\\\textbf{(rows/s)}} & 
            \makecell{\textbf{Latency}\\\textbf{Speedup}} & 
            \makecell{\textbf{Throughput}\\\textbf{Speedup}} \\ \midrule
            
            \makecell[l]{Q1: Semantic \\ Filter} & 
            \makecell{48124.34  $\pm$ 150.55} & 
            \makecell{9017.40  $\pm$ 0.09} & 
            \makecell{5508  $\pm$ 0} & 
            5.34x & 
            5.30x \\ \addlinespace
            
            \makecell[l]{Q2: Hybrid \\ Filter ($>$ 70.0)} & 
            \makecell{20590.13  $\pm$ 130.27} & 
            \makecell{9017.13  $\pm$ 0.08} & 
            \makecell{5508  $\pm$ 0} & 
            2.28x & 
            5.31x \\ \addlinespace
            
            \makecell[l]{Q3: Semantic \\ Map ($<$ 40.0)} & 
            \makecell{5344.06  $\pm$ 112.61} & 
            \makecell{11388.57  $\pm$ 0.06} & 
            \makecell{4361  $\pm$ 0} & 
            0.47x & 
            4.29x \\ \bottomrule
        \end{tabular}%
    }
\end{table}

Our unified compiled exeuction approach demonstrates strong speedups on the semantic analytics workloads. For Query 1, we achieve a strong 5.34x latency speedup and 5.30x throughput speedup over the JIT-compiled baseline on dense records (low selectivity). The throughput gains remain highly robust even for hybrid queries where a large fraction of records match the filter (2.28x latency speedup for Query 2). This gain is driven by the compiiler's ability to perform automatic multi-device hardware sharding (PMAP/GSPMD) to parallelize the LLM forward passes, which fuse and parallelize both SQL operator and AI semantic operators across available devices without manual rewriting the database and/or the ML serving engines. The baseline uses a single-process database engine on CPU, feeding data to separate TPU inference engine on TPUs; this is less efficient without extra parallelization and tuning efforts across the execution engines. In Section~\ref{sec:horizontal_scaling} we discuss how the unified program executinon can be parallelized using the compiler's auto-sharding feature, achieving linear horizontal scaling. 

It is also important to note that the baseline CPU-TPU split execution runs faster for Q3. This is because the filter selectivity is high and leaves only 10.9\% of the rows for LLM inference. This early data skipping is more advantageous running a static unified computation -- where the filtering happens as part of the unified computation on TPUs with full data. 

\subsection{Impact of Filter Selectivity}\label{sec:filter_selectivity}

The filter selectivity and how much data we are pushing to the downstream semantic operators for expensive LLM inferences governs the performance trade-offs between the baseline and the unified approach.

\begin{table}[ht]
    \centering
    \caption{Q3 performance across movie rating selectivity sweeps, \textit{values\_a < Thresh}.}
    \label{tab:selectivity_sweep_sem_ai_transform}
    \resizebox{\columnwidth}{!}{%
        \begin{tabular}{@{}lcccc@{}}
            \toprule
            \makecell[l]{\textbf{Thresh}} & 
            \makecell{\textbf{Baseline Latency}\\\textbf{(ms)}} & 
            \makecell{\textbf{Unified Latency}\\\textbf{(ms)}} & 
            \makecell{\textbf{Latency}\\\textbf{Speedup}} & 
            \makecell{\textbf{Throughput}\\\textbf{Speedup}} \\ \midrule
            
            10.0 & 
            \makecell{171.98  $\pm$ 92.64} & 
            \makecell{11383.53 $\pm$ 0.14} & 
            0.02x & 
            9.87x \\ \addlinespace
            
            30.0 & 
            \makecell{2140.67  $\pm$ 97.35} & 
            \makecell{11383.35  $\pm$ 0.06} & 
            0.19x & 
            4.35x \\ \addlinespace
            
            50.0 & 
            \makecell{11362.98  $\pm$ 108.22} & 
            \makecell{11383.32  $\pm$ 0.06} & 
            1.00x & 
            4.24x \\ \addlinespace
            
            70.0 & 
            \makecell{27350.21  $\pm$ 131.81} & 
            \makecell{11383.32  $\pm$ 0.06} & 
            2.40x & 
            4.21x \\ \addlinespace
            
            90.0 & 
            \makecell{43947.62  $\pm$ 153.13} & 
            \makecell{11383.41  $\pm$ 0.14} & 
            3.86x & 
            4.20x \\ \bottomrule
        \end{tabular}%
    }
\end{table}

We swept the filter threshold values on the semantic mapping query, Q3  (AI\_TRANSFORM), illustrated in Table~\ref{tab:selectivity_sweep_sem_ai_transform}. 
We make an interesting observation where the high filter selectivity (Thresh=10.0) favors the baseline, split execution model. That is, we can eagerly skip most of the data for LLM inference computation, whereas the whole AI query JIT compilation needs to process the entire rows with a static micro-batch size. Notice the constant latency across different selection thresholds, driven by the static program shape and the fixed number of micro-batches over 49,664 rows.

Despite the latency penalty at high selectivity, the unified compiled execution achieves throughput speed-ups (4.2x to 9.8x rows/sec) as selection fraction increases (more rows to queue for LLM inference).

\subsection{Latency Breakdown and Data Transfer Bottlenecks}
\label{sec:latency_breakdown_bottlenecks}

\begin{figure}[t]
  \centering
  \includegraphics[width=\columnwidth,height=0.6\columnwidth,trim={0cm 0cm 0cm 2cm}, clip]{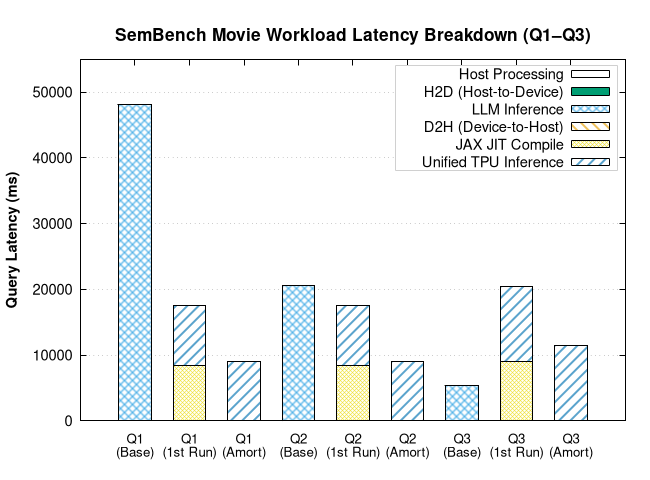}
  \caption{Latency breakdown for Queries 1-3 for the baseline (split execution) and the unified approach. Note that we compile once (1st Run) and re-use the compiled program to amortize the compilation cost across the subsequent micro-batches (Amort). }
  \label{fig:latency_breakdown}
\end{figure}

Figure~\ref{fig:latency_breakdown} illustrates the latency breakdown for Queries 1-3. Looking closely at the latency breakdown, we outline following key systems insights.

\noindent\textbf{Sequential accelerator execution dominates the baseline:} 
The overwhelming execution cost in the baseline is sequential LLM inference serving queue time on the TPU. 
The input data is streamed to TPUs as micro-batches (bsz=128) and the baseline incurs a lot of context switching overhead between CPU SQL database engine and TPU LLM inference (LLM Inference >> Unified TPU Inference). The unified JAX program, on the other hand, either loads the entire raw tabular data once or process the micro-batches without the expensive CPU-TPU synchronization. The compiler can also fuse SQL operators with LLM layers (e.g., scalar operations with embedding lookups or initial tokenization steps).

\begin{figure}[t]
  \centering
  \begin{subfigure}[b]{0.49\columnwidth}
    \centering
    \includegraphics[width=\textwidth, height=3.2cm, trim={0cm 0cm 0cm 1.5cm}, clip]{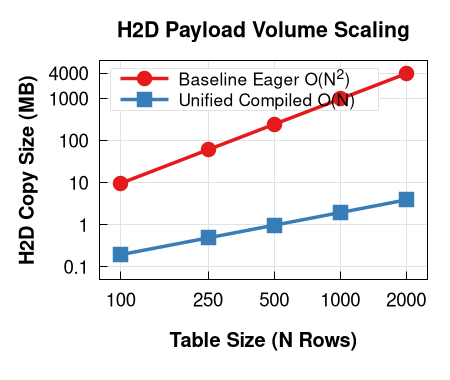}
    \caption{Payload volume scaling}
    \label{fig:cross_product_payload}
  \end{subfigure}
  \hfill 
  \begin{subfigure}[b]{0.495\columnwidth}
    \centering
    \includegraphics[width=\textwidth, height=3.2cm, trim={0cm 0cm 0cm 1.5cm}, clip]{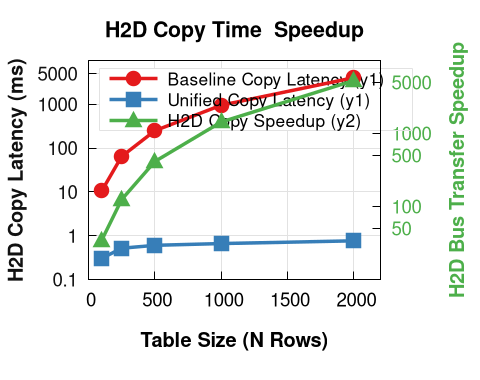}
    \caption{Copy time speedups}
    \label{fig:cross_product_latency}
  \end{subfigure}
  
  \caption{The cartesian (cross-join) Host-to-Device (H2D) data transfer overheads and speedups for Q4 execution}
  \label{fig:cross_product_query}
\end{figure}

\noindent\textbf{Data transfers are not primary latency bottlenecks for simple aggregation and transformation queries.} 
The total data transfer overhead is negligible (0.2\% of the baseline execution) for the single  relation data that fits locally and with compact output formats (e.g., Yes/No classification results).
While negligible, this PCIe transfer bottleneck can escalate dramatically with queries that require a large data transfer (H2D) and/or intermediate token embeddings or logits back to the host (D2H) for confidence-based soft filtering and even debugging. For instance, the cartesian (cross-join) query, Q4 from Table~\ref{tab:ai_sql_queries} for identifying almost identical movie reviews would requires sending $O(N^2)$ (N: relation size) input rows/tuples to the ML serving engine for the baseline, whereas only $O(N)$ for the unified approach.
As illustrated in Figure~\ref{fig:cross_product_query}, the cartesian H2D data transfer quickly becomes a system bottleneck; even for a small table size (N=2000), the transfer copy latency escalates to 4.06 seconds. By contrast, the dynamic data movement footprint remains strictly linear with the unified compiled approach.  Rather than materializing duplicating review pairs on the host, the unified program requires a negligible 3.9 MB copy transfer size vs. 3.9 GB at N=2000.


\begin{figure}[t]
  \centering
  \begin{subfigure}[b]{0.49\columnwidth}
    \centering
    \includegraphics[width=\textwidth, height=3.2cm, trim={0cm 1cm 0cm 1.5cm}, clip]{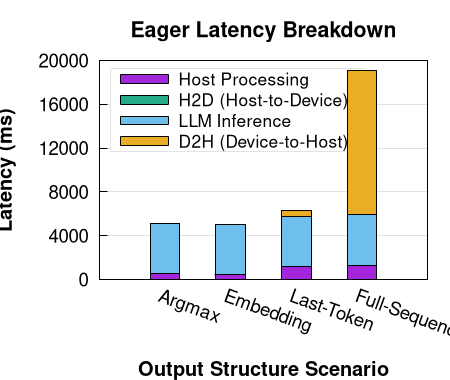}
    \caption{Baseline latency breakdown}
    \label{fig:data_format_payload}
  \end{subfigure}
  \hfill 
  \begin{subfigure}[b]{0.495\columnwidth}
    \centering
    \includegraphics[width=\textwidth, height=3.2cm, trim={0cm 1cm 0cm 1.5cm}, clip]{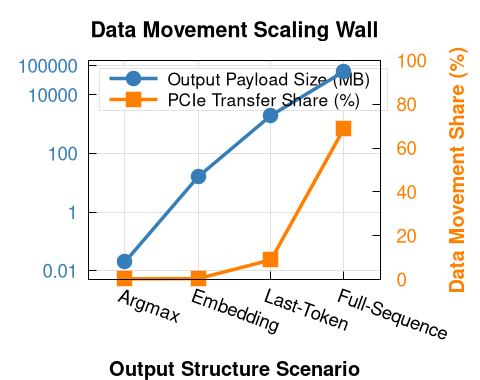}
    \caption{Data movement scaling wall}
    \label{fig:data_format_scaling}
  \end{subfigure}
  
  \caption{Device-to-Host (D2H) data transfer overheads and scaling by output format/structure for Q5 execution}
  \label{fig:data_movement_wall}
\end{figure}

LLM inference results can also potentially escalate the data transfer (D2H) bottleneck. The simple test queries Q1-3, returned argmax output tokens (e.g., Yes, No binary classification results) only; however, for analytics cases where raw distributions or logits are required for debugging/explaining the semantic operator decisions, the output payload size, thus the D2H data transfer latency can grow significantly, as shown in Figure~\ref{fig:data_movement_wall}. 
Query 5 from Table~\ref{tab:ai_sql_queries} aggregates over long sequence outputs that contain a word \texttt{\%masterpiece\%}. As with the sequence length, the output formats (e.g., argmax tokens, latent vectors/embeddings, logits for the last token, fulll sequence logits) change the size of the payload drastically. The key challenge here is the host-side \textit{LIKE} string match, that require moving all the transformation results back to the host for filtering. For the unified execution, we fuse and execute the string matching directly on local HBM, transferring the debug outputs for the matching essays only. Even without the extra debug outputs, the unified execution only needs to transfer the aggregation result, whereas the baseline still needs everything for the host processing of the aggregation.

\subsection{Horizontal Scaling}\label{sec:horizontal_scaling}

This unified approach provides more than just the data movement latency elimination, unlocking new opportunities for global cross-boundary optimizations for AI queries. 
A major global optimization enabled by this unified program is unified, compiler-driven multi-device sharding. In traditional split AI Query Engine architectures, developers must manually write, calibrate, and synchronize two completely independent parallelization strategies: one for the CPU database engine (e.g., dynamic core-level partitioning) and another for the TPU/GPU machine learning engine (e.g., tensor-parallel or pipeline-parallel model sharding). This requires complex dynamic load balancing across the CPU-TPU boundary. By contrast, because our compiler translates both relational database scans and LLM inferences into a single unified tensor program, the compiler can globally shard both the relational data arrays and the neural computations concurrently and automatically via XLA's GSPMD/PMAP. 
This is very convenient and also very effective, as we illustrate in Table~\ref{tab:horizontal_scaling}. We evaluated the horizontal scaling capability of our system by sweeping the number of TPU devices from 1 to 4.

\begin{table}[ht]
    \centering
    \caption{Horizontal scaling performance across multiple TPU devices using Q3 with \texttt{values\_a < 50.0}}
    \label{tab:horizontal_scaling}
    \resizebox{\columnwidth}{!}{%
        \begin{tabular}{@{}lccc@{}}
            \toprule
            \textbf{Devices} & \textbf{TPU Time (ms)} & \textbf{Throughput (rows/s)} & \textbf{Speedup} \\ \midrule
            1 & $45753.39 \pm 0.04$ & $1091 \pm 0$ & 1.00x \\ \addlinespace
            2 & $22878.20 \pm 0.03$ & $2182 \pm 0$ & 2.00x \\ \addlinespace
            4 & $11383.98 \pm 0.07$ & $4363 \pm 0$ & 4.00x \\ \bottomrule
        \end{tabular}%
    }
\end{table}

We replicate the same unified compiled program on all devices, but with different micro-batches across them. This enables near-perfect linear horizontal scaling, per the simple Aggregation example query. This demonstrates that the unified compiled model is extremely scalable, just like any LLM workloads with data parallelism, and fully prepared for massive distributed execution workloads. 
It is important to note that our preliminary evaluation looked at intra-node multi-device scaling. This overlooks the large-scale sharding as well as inter-node network overheads. We plan to investigate the multi-node scaling in the future.

\subsection{Multi-Tenancy \& Concurrent Query Bottlenecks}

In real-world enterprise database systems, multiple users (tenants) execute queries concurrently. While specialized LLM serving engines (e.g., vLLM) employ highly engineered continuous batching and tensor parallelism to handle concurrent requests from multiple clients, our system compiles each query as a standalone JAX program/executable.

\begin{table}[ht]
    \centering
    \caption{Multi-tenant concurrent execution scalability.}
    \label{tab:multitenant_scalability}
    \resizebox{\columnwidth}{!}{%
        \begin{tabular}{@{}lccc@{}}
            \toprule
            \makecell[l]{\textbf{Concurrent Tenants} \\ \textbf{($K$)}} & 
            \makecell{\textbf{Average Latency}\\\textbf{(ms)}} & 
            \makecell{\textbf{Aggregate Throughput}\\\textbf{(rows/s)}} & 
            \makecell{\textbf{Throughput}\\\textbf{Scaling}} \\ \midrule
            
            1 & 
            \makecell{3624.33  $\pm$ 0.03} & 
            \makecell{1376  $\pm$ 0} & 
            1.00x \\ \addlinespace
            
            2 & 
            \makecell{5435.86  $\pm$ 0.01} & 
            \makecell{1377  $\pm$ 0} & 
            1.00x \\ \addlinespace
            
            4 & 
            \makecell{9058.71  $\pm$ 0.14} & 
            \makecell{1377  $\pm$ 0} & 
            1.00x \\ \addlinespace
            
            6 & 
            \makecell{12681.62  $\pm$ 0.17} & 
            \makecell{1377  $\pm$ 0} & 
            1.00x \\ \addlinespace
            
            8 & 
            \makecell{16303.90  $\pm$ 0.41} & 
            \makecell{1377  $\pm$ 0} & 
            1.00x \\ \bottomrule
        \end{tabular}%
    }
\end{table}

To evaluate the scalability of compiled AI queries under multi-tenant workloads, we spawned $k$ concurrent client threads, each submitting an instance of the compiled \textit{AI\_IF} semantic filter query, where each query has to process ~5000 rows. All threads shared the same on-device \textit{Gemma-2B} parameters and invoked the same JIT-compiled program concurrently. We swept $k$ from 1 to 8 and measured aggregate throughput (total rows processed across all clients divided by total elapsed time) and individual client latencies. Table~\ref{tab:multitenant_scalability} summarizes the results.

The experimental results reveal a bottleneck in concurrent execution: the aggregated throughput remains constant at exactly 1377 rows/sec, while average query latency scales linearly with the number of tenants. This is because the TPU execution queue is serialized and the JAX programs from different and concurrent queries must run one after the other (FIFO queue). 
To support the unified compilation paradigm in high-throughput enterprise DBMS, we need to support a runtime scheduler that can dynamically schedule and re-schedule different programs in the TPU queue to meet the SLO and increase the throughput, like continuous batching in vLLM.

\section{Conclusion \& Research Roadmap}
In this vision paper, we demonstrated the feasibility and potential performance gains of compiling hybrid AI queries as a whole into a single, unified JAX tensor program for execution on specialized accelerators (TPUs). By consolidating accelerator boundaries and compiling a single unified program, we successfully amortize JAX JIT compilation overheads and enable compiler-driven multi-device sharding and optimizations for the CPU databse and TPU ML inference engines (up to 4.00x scaling on 4 devices) on both synthetic benchmarks and \textit{SemBench} workloads. To realize the full potential of this unified execution strategy, several key research challenges must be solved by the systems and database community:

\noindent\textbf{Semantic-Aware Cost-Based Query Optimization:} Traditional database optimizers rely on relational cost models (e.g., cardinality, selectivity) to choose execution plans. For the modern AI Query Engine, the optimizer must balance relational selectivity with LLM-based AI semantic operator costs. 
This cost-based AI Query optimization pose a similar challenge as the inlining vs. outlining trade-offs for traditional UDF optimizer~\cite{arch2026partial}. Instead of deciding whether to inline procedural CPU-based UDFs, an AI Query optimizer may choose between the unified compiled execution and the split execution strategies.

\noindent\textbf{Dynamic Shape Compilation:} JAX/XLA compilers rely heavily on static tensor shapes to genreate highly optimized TPU/GPU machine executables, and to avoid frequent re-compilation. However, real-wokrld AI queries can involve input data of highly varying lengths/sizes, requiring dynamic input shapes. The current prototype system pad inputs to static maximum sequence lengths, which waste TPU memory and compute. Research is needed to enable efficient dynamic-shape tracing for the hybird AI query workloads.

\noindent\textbf{Multi-Query Concurrent Execution:} In high-throughput enterprise databases, multiple queries run concurrently. The current stack highly optimizes for static graph and shape specialization, taking one executable at a time. This limits the performance with multi-tenancy, where more dynamic scheduling and execution, possibly with token interleaving across independent concurrently queries are desired~\cite{yu2022orca}.

\noindent\textbf{Multi-Modal Compilation:} Future AI qureis will involve multi-modal oeprators processing text, image, audio, video, etc. More work is needed to support a wide range of operators for efficient compilation, but also the increasing sizes and complex model achitectures raise  sharding and partitioning challenges.

\noindent\textbf{Dynamic TPU Data Skipping:} One of the challenges in the unified program JIT compilation is the lack of dynamic, on-device data skipping. As discussed in Section~\ref{sec:filter_selectivity}, the baseline, split execution outperforms the unified executnion with high selectivity filter predicates. The static tracing and compilation forces full table scan regardless selectivity. Research is needed to solve this compile vs. skip trade-off.

\begin{acks}
We would like to thank Hank Levy, David Culler, Greg Ganger, Arvind Krishnamurthy and Brian Suchy for their valuable feedback and discussion.
\end{acks}


\bibliographystyle{ACM-Reference-Format}
\bibliography{main}

\end{document}